\documentclass{moriond}
\usepackage{mathtools}

\def\Journal#1#2#3#4{{#1} {\bf #2}, #3 (#4)}

\def\PRL{\em Phys. Rev. Lett.}
\def\PRD{{\em Phys. Rev.} D}

\begin{document}
\title{FIRST DETECTION OF CNO NEUTRINOS WITH BOREXINO}

\author{ G. SETTANTA$^1$, on behalf of the Borexino Collaboration:\\
M. Agostini$^{2,3}$, K. Altenm\"{u}ller$^3$, S. Appel$^3$, V. Atroshchenko$^4$, Z. Bagdasarian$^{1,\,}$\thanks{Berkeley}, D. Basilico$^5$, G. Bellini$^5$, J.Benziger$^6$, R. Biondi$^7$, D. Bravo$^{5,\,}$\thanks{Madrid}, B. Caccianiga$^5$, F. Calaprice$^8$, A. Caminata$^{9}$, P. Cavalcante$^{10,\,}$\thanks{LNGSG}, A. Chepurnov$^{11}$, D. D’Angelo$^5$, S. Davini$^{9}$, A. Derbin$^{12}$, A. Di Giacinto$^7$, V. Di Marcello$^7$, X.F. Ding$^8$, A. Di Ludovico$^8$, L. Di Noto$^{9}$, I. Drachnev$^{12}$, A. Formozov$^{13,6}$, D. Franco$^{14}$, C. Galbiati$^{8,15}$, C. Ghiano$^7$, M. Giammarchi$^5$, A. Goretti$^{8,\,c}$, A. S. G\"{o}ttel$^{1,16}$, M. Gromov$^{11,14}$, D. Guffanti$^{17}$, Aldo Ianni$^7$, Andrea Ianni$^8$, A. Jany$^{18}$, D. Jeschke$^3$, V. Kobychev$^{19}$, G.Korga$^{20,21}$, S. Kumaran$^{1,16}$, M. Laubenstein$^7$, E. Litvinovich$^{4,22}$, P. Lombardi$^5$, I. Lomskaya$^{12}$, L. Ludhova$^{1,16}$, G. Lukyanchenko$^4$, L. Lukyanchenko$^4$, I. Machulin$^{4,22}$, J. Martyn$^{17}$, E. Meroni$^5$, M. Meyer$^{23}$, L. Miramonti$^5$, M. Misiaszek$^{18}$, V. Muratova$^{12}$, B. Neumair$^3$, M. Nieslony$^{17}$, R. Nugmanov$^{4,22}$, L. Oberauer$^3$, V. Orekhov$^{17}$, F. Ortica$^{24}$, M. Pallavicini$^{9}$, L. Papp$^3$, L. Pelicci$^{1,16}$, \"{O}. Penek$^{1,16}$, L. Pietrofaccia$^8$, N. Pilipenko$^{12}$, A. Pocar$^{25}$, G. Raikov$^4$, M. T. Ranalli$^7$, G. Ranucci$^5$, A. Razeto$^7$, A. Re$^5$, M. Redchuk$^{1,16,\,}$\thanks{Padova}, A. Romani$^{24}$, N. Rossi$^7$, S. Sch\"{o}nert$^3$, D. Semenov$^{12}$, M. Skorokhvatov$^{4,22}$, A. Singhal$^{1,16}$, O. Smirnov$^{13}$, A. Sotnikov$^{13}$, Y. Suvorov$^{4,8,\,}$\thanks{Napoli}, R. Tartaglia$^7$, G. Testera$^{9}$, J. Thurn$^{23}$, E. Unzhakov$^{12}$, F. L. Villante$^{7,26}$, A. Vishneva$^{13}$, R. B. Vogelaar$^{10}$, F. von Feilitzsch$^3$, M. Wojcik$^{18}$, M. Wurm$^{17}$, S. Zavatarelli$^{9}$, K. Zuber$^{23}$, G. Zuzel$^{18}$.}

\vspace{1cm}

\address{$^{1}$Institut f\"ur Kernphysik, Forschungszentrum J\"ulich, 52425 J\"ulich, Germany}
\address{$^{2}$Department of Physics and Astronomy, University College London, London, UK}
\address{$^{3}$Physik-Department, Technische Universit\"at  M\"unchen, 85748 Garching, Germany}
\address{$^{4}$National Research Centre Kurchatov Institute, 123182 Moscow, Russia}
\address{$^{5}$Dipartimento di Fisica, Universit\`a degli Studi e INFN, 20133 Milano, Italy}
\address{$^{6}$Chemical Engineering Department, Princeton University, Princeton, NJ 08544, USA}
\address{$^{7}$INFN Laboratori Nazionali del Gran Sasso, 67010 Assergi (AQ), Italy}
\address{$^{8}$Physics Department, Princeton University, Princeton, NJ 08544, USA}
\address{$^{9}$Dipartimento di Fisica, Universit\`a degli Studi e INFN, 16146 Genova, Italy}
\address{$^{10}$Physics Department, Virginia Polytechnic Institute and State University, Blacksburg, VA 24061, USA}
\address{$^{11}$Lomonosov Moscow State University Skobeltsyn Institute of Nuclear Physics, 119234 Moscow, Russia}
\address{$^{12}$St. Petersburg Nuclear Physics Institute NRC Kurchatov Institute, 188350 Gatchina, Russia}
\address{$^{13}$Joint Institute for Nuclear Research, 141980 Dubna, Russia}
\address{$^{14}$AstroParticule et Cosmologie, Universit\'e Paris Diderot, CNRS/IN2P3, CEA/IRFU, Observatoire de Paris, Sorbonne Paris Cit\'e, 75205 Paris Cedex 13, France}
\address{$^{15}$Gran Sasso Science Institute, 67100 L'Aquila, Italy}
\address{$^{16}$RWTH Aachen University, 52062 Aachen, Germany}
\address{$^{17}$Institute of Physics and Excellence Cluster PRISMA+, Johannes Gutenberg-Universit\"at Mainz, 55099 Mainz, Germany}
\address{$^{18}$M.~Smoluchowski Institute of Physics, Jagiellonian University, 30348 Krakow, Poland}
\address{$^{19}$Kiev Institute for Nuclear Research, 03680 Kiev, Ukraine}
\address{$^{20}$Department of Physics, Royal Holloway, University of London, Department of Physics, School of Engineering, Physical and
Mathematical Sciences, Egham, Surrey, TW20 OEX, UK}
\address{$^{21}$Institute of Nuclear Research (Atomki), Debrecen, Hungary}
\address{$^{22}$ National Research Nuclear University MEPhI (Moscow Engineering Physics Institute), 115409 Moscow, Russia}
\address{$^{23}$Department of Physics, Technische Universit\"at Dresden, 01062 Dresden, Germany}
\address{$^{24}$Dipartimento di Chimica, Biologia e Biotecnologie, Universit\`a degli Studi e INFN, 06123 Perugia, Italy}
\address{$^{25}$Amherst Center for Fundamental Interactions and Physics Department, University of Massachusetts, Amherst, MA 01003, USA}
\address{$^{26}$Dipartimento di Scienze Fisiche e Chimiche, Universit\`a dell'Aquila, 67100 L'Aquila, Italy}

\address{$^a$Present address: University of California, Berkeley, Department of Physics, CA 94720, Berkeley, USA}
\address{$^b$Present address: Universidad Autónoma de Madrid, Ciudad Universitaria de Cantoblanco, 28049 Madrid, Spain}
\address{$^c$Present address: INFN Laboratori Nazionali del Gran Sasso, 67010 Assergi (AQ), Italy}
\address{$^d$Present address: Dipartimento di Fisica e Astronomia dell'Università di Padova and INFN Sezione di
Padova, Padova, Italy}
\address{$^e$Present address: Dipartimento di Fisica, Universit\`a degli Studi Federico II e INFN, 80126 Napoli, Italy}

\vspace{1cm}
\maketitle
\abstracts{
Neutrinos are elementary particles which are known since many years as fundamental messengers from the interior of the Sun. The Standard Solar Model, which gives a theoretical description of all nuclear processes which happen in our star, predicts that roughly 99\% of the energy produced is coming from a series of processes known as the “\emph{pp} chain”. Such processes have been studied in detail over the last years by means of neutrinos, thanks also to the important measurements provided by the Borexino experiment. The remaining 1\% is instead predicted to come from a separate loop-process, known as the “CNO cycle”. This sub-dominant process is theoretically well understood, but has so far escaped any direct observation. Another fundamental aspect is that the CNO cycle is indeed the main nuclear engine in stars more massive than the Sun. In 2020, thanks to the unprecedented radio-purity and temperature control achieved by the Borexino detector over recent years, the first ever detection of neutrinos from the CNO cycle has been finally announced. The milestone result confirms the existence of this nuclear fusion process in our Universe.  Here, the details of the detector stabilization and the analysis techniques adopted are reported.
}

\section{Solar physics and CNO neutrinos}

\begin{figure}[t]
\begin{center}
    \includegraphics[width=0.7\textwidth]{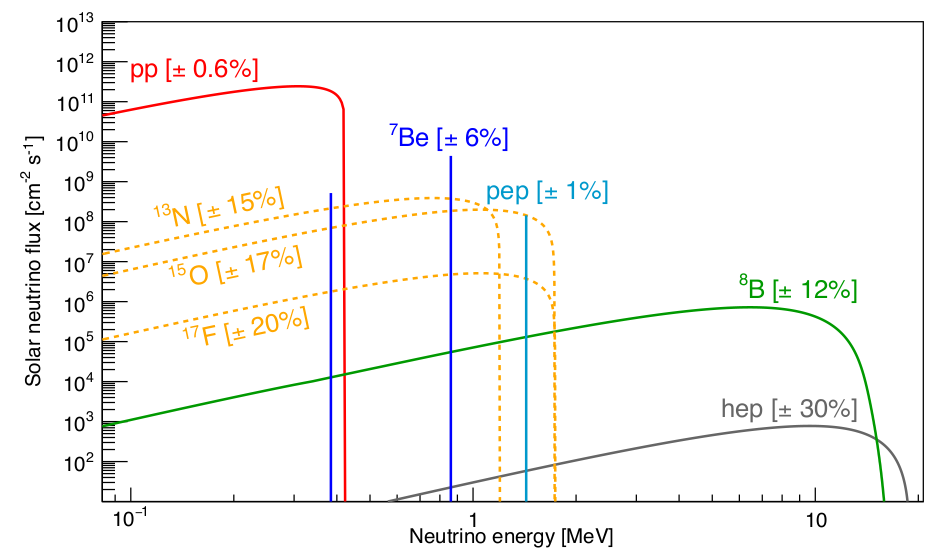}
    \caption{Solar-neutrinos energy spectra obtained from \url{http://www.sns.ias.edu/~jnb/}, using fluxes taken from~\protect\cite{B16SSM}. The \emph{pp} (red), \emph{pep} (light blue), $^7$Be (blue), $^8$B (green), \emph{hep} (grey) and CNO (orange) neutrino fluxes are shown separately.}
     \label{fig:nu_spectra}
\end{center}
\end{figure}

Solar neutrinos are produced as the result of nuclear fusion processes that occur in the core of the Sun, in the electron flavor ($\nu_e$). Differently from light, they are able to travel from the production site until the Earth, without being deflected or absorbed, and are therefore a direct probe to the Sun’s interior. They have been indeed extensively used in the past to understand the fundamental processes that power our star. From them, it was also obtained the experimental evidence of flavor oscillation in the neutrino sector~\cite{SNO_1,SNO_2}.\\
Predictions of solar neutrino fluxes are provided by the \emph{Standard Solar Model} (SSM), which summarizes the current best knowledge about the physics of the Sun, and of stars in general. The vast majority of the energy produced in the Sun comes from a series of reactions called \emph{pp} chain, which accounts for about 99\% of the solar luminosity. The associated neutrino flux is generated by various sub-processes, like the \emph{pp} and the $^7$Be reactions.\\
Beside the \emph{pp} chain, another process is able to produce heat in the Sun’s core. Only recently detected for the first time, the CNO cycle accounts for the remaining 1\% in the solar luminosity balance. The net contribution of the cycle is strongly dependent upon the core’s temperature and metal content, which is why it is believed to be the dominant mechanism for energy conversion in heavier stars~\cite{Solar_book05}. Neutrinos produced in the CNO cycle contribute to the mid-energy range, partially overlapping with the \emph{pep} spectrum. Figure \ref{fig:nu_spectra} reports the energy spectra of solar neutrinos, as produced by the different processes.\\
The solar neutrino flux provides a direct information of solar fusion processes. Although neutrinos small cross section is not influencing their own trajectory, nor their energy, they are still subject to flavor conversion effects. Due to the dense environment of the Sun and the distance traveled, the flux detected at Earth is no longer 100\% made of $\nu_e$, as it was at production. 

\noindent Neutrinos produced in the CNO cycle are a unique probe to the Sun’s primordial composition. The CNO contribution to the Sun’s fusion processes depends on the core’s metallicity, i.e. the mass abundance of elements heavier than helium. The Sun is a relatively small and cold star in the Universe and the CNO contribution to the luminosity budget is around 1\%. For heavier stars, having a mass greater than $\sim$1.3 solar masses, the CNO cycle is instead believed to be the dominant process which burns hydrogen into helium. It is therefore considered as the main nuclear fusion process occurring in the Universe.  The precise CNO contribution in the Sun, however, is unknown. The reason is that the Sun’s metallicity is not known with precision. The metallicity indeed influences directly the efficiency of the cycle, since the “metals” Carbon, Nitrogen and Oxygen are the elements who catalyze the process. The CNO neutrino flux is then directly dependent upon the core’s metallicity, which keeps memory of the Sun’s elemental composition at the time of formation. Since the metal abundance in the core is believed not to be influenced by the surface, CNO neutrinos preserve the core’s information at the pristine conditions.\\
Within the SSM, one can distinguish among the Low-metallicity (LZ) and High-metallicity (HZ) scenarios, the first one based on three-dimensional hydro-dynamical models of the solar atmosphere~\cite{LZ1,LZ2} and the second one on a older one-dimensional modeling of the solar atmosphere~\cite{HZ}. LZ model predicts a lower metal abundace in the Sun's core than the HZ one and the ambiguity is known as \emph{metallicity puzzle}. This ambiguity is also enhanced from the fact that Helioseismology measurements seem to prefer the HZ scenarios~\cite{B16SSM}, while recent observations of Sun's surface luminosity point towards LZ scenarios. A precise measurement of the CNO neutrino flux could be the solution to the metallicity puzzle, since the flux prediction changes as much as $\sim$30\% in case of LZ or HZ scenario assumption.

\section{The Borexino detector}
\begin{figure}[t]
\begin{center}
    \includegraphics[width=0.6\textwidth]{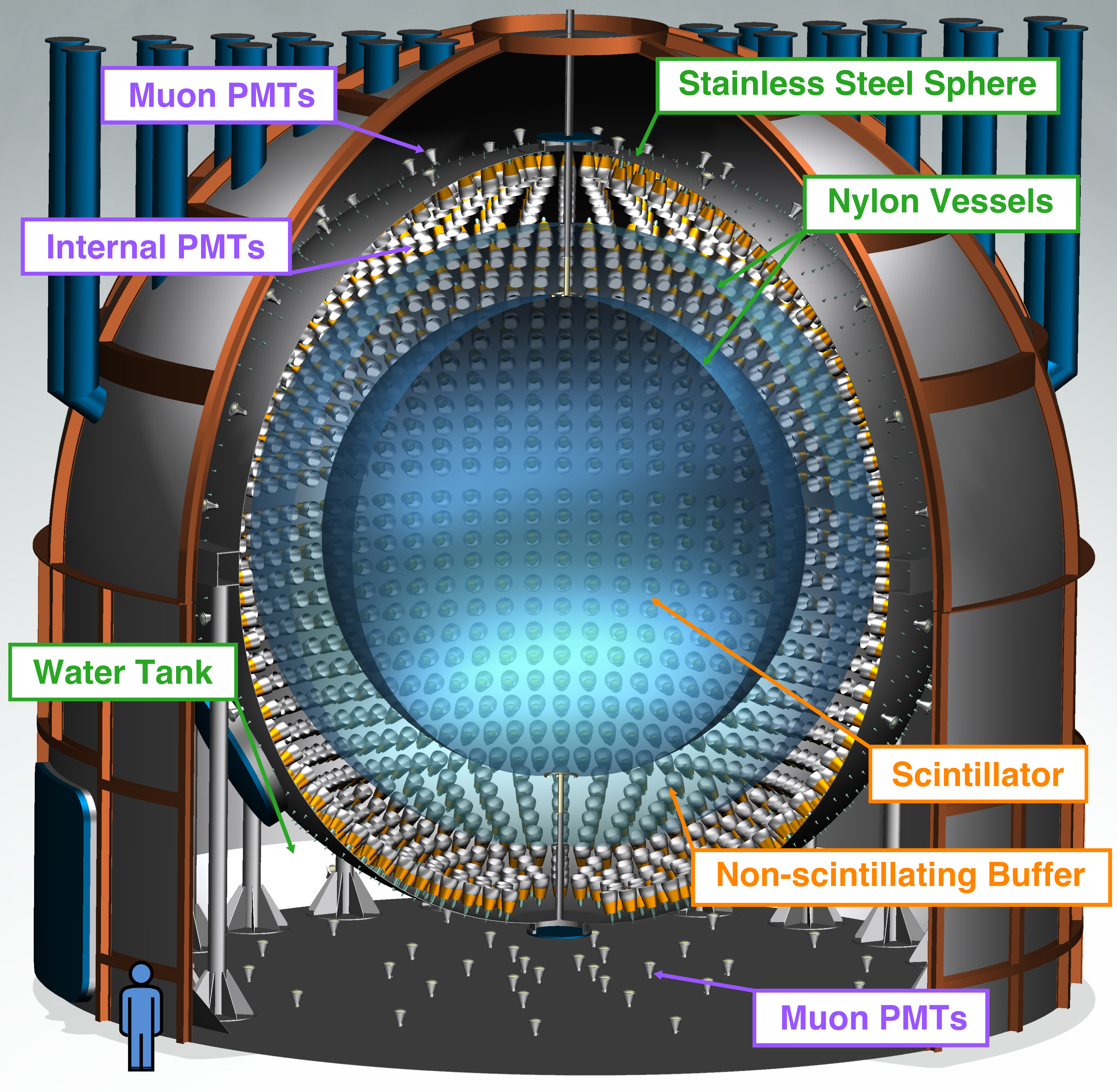}
    \caption{Sketch of the Borexino detector.}
     \label{fig:bx_detector}
\end{center}
\end{figure}

Borexino is a large, low-background, liquid scintillator-based neutrino detector, located at the Laboratori Nazionali del Gran Sasso (Italy). The particular location in the hearth of the Gran Sasso Massif, under $\sim$3.800 m.w.e., ensure that the Cosic Ray-induced rate of muons is reduced by a factor $\sim10^6$. The detector consists of a layered structure. The innermost part is the liquid scintillator (LS) volume contained inside a nylon Inner Vessel (IV), which makes the primary target for solar neutrinos. The radius of the IV is 4.25\,m. The light produced by the interactions of neutrinos with the LS is collected by $\sim$2000 Photo-multiplier Tubes (PMTs), installed on a spherical steel structure with a radius of 6.85\,m. To reduce the penetration of gamma radiation coming from the steel and the PMTs into the LS, the space between the PMTs and the IV is filled with a non-scintillating buffer liquid and another nylon vessel is placed in between. The whole spherical structure is then contained inside an outer water tank, equipped with $\sim$200 PMTs, which is used as an active veto for cosmic radiation. A sketch of the detector is in Figure \ref{fig:bx_detector}.\\
The detection principle is based on neutrino-electron elastic scattering and therefore Borexino is sensitive to all neutrino families. However, since the cross-section of charged-current (CC) interactions is higher by a factor $\sim$6 with respect to neutral-current (NC) interactions, Borexino essentially measures only electron neutrinos, coming from the Sun.

\noindent Borexino data-taking started in 2007 and has seen three different phases among its lifetime. During Phase-I, from 2007 and 2010, the experiment released the first precision measurement of $^7$Be neutrinos~\cite{Be7} and of their flux modulation~\cite{LP_Ph1}, together with the first direct observation of \emph{pep} neutrinos~\cite{pep}. At the end of Phase-I the detector underwent a LS purification campaign, which reduced the presence of several contaminants, like $^{85}$Kr and $^{210}$Bi. After that, at the end of 2011, the Phase-II data-taking period started. With the data collected during the whole Phase-II, until May 2016, Borexino was able to provide a comprehensive measurement of all the \emph{pp} chain neutrino components~\cite{Nat18,PRD19}. Thanks to the refined measurement of $^7$Be and $^8$B fluxes, which are also sensitive to the Sun's metallicity, it was possible to extract a mild preference for the HZ scenario, with a significance around 2$\sigma$~\cite{Nat18}. In an effort to stabilize the detector, to address the detection of CNO neutrinos, the Borexino Collaboration completed the detector thermal insulation between 2015 and 2016. The latest data-acquisition period of Borexino (Phase-III) started in July 2016 and saw an improved thermal stability of the detector.

\section{The CNO analysis}
The energy of recoiling electrons after a CNO neutrino interaction shows a continuous distribution with an endpoint at 1.517\,keV. The same energy region is populated by other backgrounds, which limit the sensitivity towards the CNO neutrino events. The most important ones are the $\beta^-$ particles emitted by the $^{210}$Bi contaminant and the $pep$-neutrino recoil electrons. These two species have spectral features very similar to that of CNO neutrinos and their absolute rate must then be constrained by means of independent inputs. The high-energy tail of the CNO spectrum is also contaminated by $\beta^+$ decays of the cosmogenic $^{11}$C isotope, which is treated with a dedicated tagging algorithm.\\
The rate of $pep$-neutrinos interacting in Borexino can be efficiently constrained by means of a combination of robust theoretical assumptions and various available data, down to 1.4\% precision \cite{Bergstrom_2016,CNOsens}. The other main background, consisting of $^{210}$Bi decays, has an overall rate in the detector which can be assumed to be in secular equilibrium with his parent nucleus $^{210}$Pb decay rate, given the short $^{210}$Bi lifetime. The decay chain of $^{210}$Pb can be summarized as follows:

\begin{equation} \label{eq:Pb_chain}
    { }^{210} \mathrm{Pb} \underset{22.3 \mathrm{yr}}{\stackrel{\beta^{-}}{\longrightarrow}} { }^{210} \mathrm{Bi} \underset{5 \mathrm{d}}{\stackrel{\beta^{-}}{\longrightarrow}} { }^{210} \mathrm{Po} \underset{138.4 \mathrm{d}}{\stackrel{\alpha}{\longrightarrow}} { }^{206} \mathrm{Pb}.
\end{equation}

\noindent $^{210}$Pb decays are well below the analysis threshold and can be considered therefore invisible. The $^{210}$Po, on the other hand, is an unstable isotope which produces mono-energetic $\alpha$ particles, easily detectable in Borexino on an event-by-event basis. Provided that the secular equilibrium in Equation \ref{eq:Pb_chain} is maintained, the measured $^{210}$Po corresponds to the $^{210}$Bi rate.\\

\subsection{Low Polonium Field}
As discussed previously, the rate of $^{210}$Bi decays can be constrained via its link with the $^{210}$Po decay rate, in the assumption that this latter term is only supported by in-equilibrium $^{210}$Pb decay chain. Data collected by Borexino since its start, however, indicate that an out-of-equilibrium component of $^{210}$Po is present in the detector. The source of this component is likely the surface of the IV, from which $^{210}$Po is detached into the scintillator. The mean free path of $^{210}$Po atoms is calculated to be very small in stable conditions. The presence, however, of convective motions in the Borexino scintillator allow $^{210}$Po to spread among all the scintillator volume. Under this conditions, the measured value of $^{210}$Po decay rate would be much higher than the $^{210}$Bi decay rate, spoiling any possible constraint.\\
To limit convective motions in the scintillator volume, the Borexino Collaboration pursued a careful-planned effort, culminated in the detector thermal insulation in 2015--2016 and the subsequent installation of active temperature controls. This way, $^{210}$Po mixing has been strongly suppressed since 2016, bringing to the formation of a very clean region around the centre of the detector, named the \emph{Low Polonium Field} (LPoF). The in-equilibrium $^{210}$Po decay rate in the LPoF region can be then measured. In a conservative approach, some residual contribution of convective $^{210}$Po has been considered and the measured $^{210}$Po rate has been therefore translated into an upper limit for the $^{210}$Bi rate, which resulted to be 11.5$\pm$1.3\,counts--per--day (cpd)/100\,ton. The quoted uncertainty includes both the statistical one and systematics contributions, such as the uncertainties of the $^{210}$Po rate evaluation and the $^{210}$Bi inhomogeneity.

\subsection{Spectral analysis}\label{sec:CNO_spec_ana}
To extract the CNO neutrino interaction rate from data, a multivariate analysis has been performed, simultaneously fitting the energy spectrum between 320\,keV and 2.640\,keV and the radial distribution of events, after a selection procedure which removes noise events. The data acquired between July 2016 and February 2020 have been considered for the analysis and events have been selected inside a fiducial volume of 71.3\,ton. Apart from the constrained $pep$-neutrino and $^{210}$Bi rates, all other species rates (including CNO neutrinos) have been left free to vary in the fit. The reference probability density functions (PDFs) used to fit both the energy and radial distributions have been built by means of a complete Monte Carlo simulation. Considering only statistical uncertainty, the best fit returns an interaction rate of CNO neutrinos in Borexino $R = 7.2^{+2.9}_{-1.7}$ cpd/100\,ton (68\% confidence interval)~\cite{CNOpap}.\\
The effect of the fit configuration has been found negligible in the analysis. A careful accounting of several systematic uncertainty sources in the final measurement has been performed, leading to a total systematic contribution of [-0.5, 0.6]\,cpd/100\,ton. The dominant contribution comes from the scintillator light yield uncertainty.

\subsection{Counting analysis}\label{sec:CNO_counting}
As a cross-check of the spectral analysis, a counting analysis has been performed on the same data sample. Data events have been counted inside an energy region where the CNO signal-to-background ratio is maximized, corresponding to [780 - 885] keV. The $pep$-neutrino and $^{210}$Bi rate constraints used in the spectral analysis have been used in the counting analysis as well. The CNO interaction rate is extracted by subtracting all background contributions, evaluated within a certain uncertainty, and then propagating those uncertainties. As for the spectral analysis, the dominant contribution is the uncertainty on the scintillator light yield.\\
The final measured rate is 5.6$\pm$1.6 cpd/100\,ton, confirming the presence of CNO at 3.5$\sigma$ level.

\subsection{Results}\label{sec:CNO_results}
\begin{figure}[t]
    \includegraphics[width=0.9\textwidth]{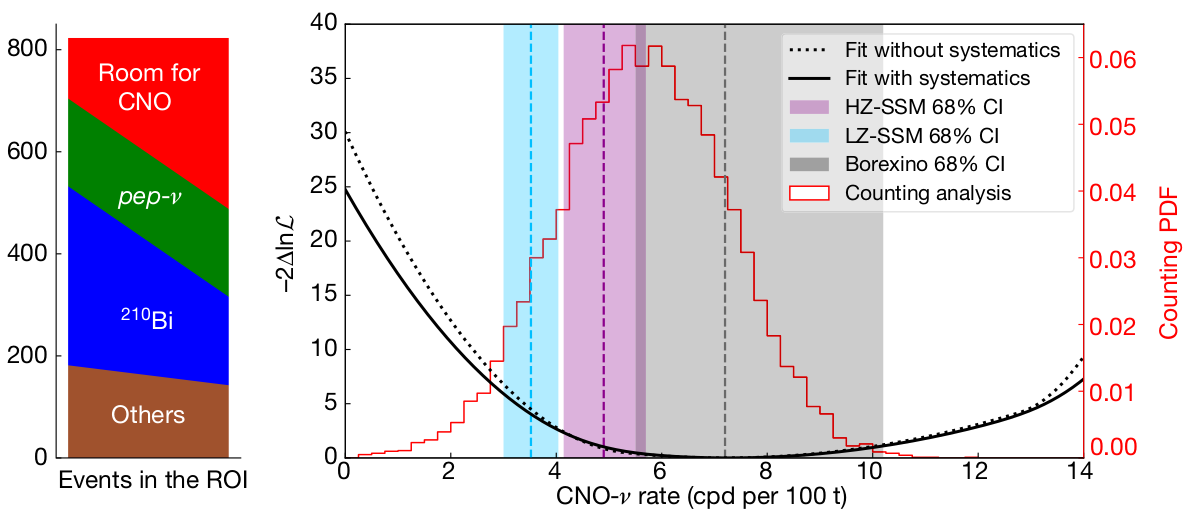}
    \caption{Summary of CNO counting and spectral analysis results. Left, counting analysis bar chart. Right, CNO-neutrino rate negative log-likelihood (ln$\mathcal{L}$) profiles, together with PDF from counting analysis and models predictions. Figure from \protect\cite{CNOpap}.}
    \label{fig:cno-final}
    \end{figure}

Figure \ref{fig:cno-final} summarizes the Borexino result on the measured CNO neutrino interaction rate. The result of the spectral fit is reported in terms of a log-likelihood profile, with statistical uncertainty only and also folded with systematics contributions. The best fit value is then $R = 7.2^{+3.0}_{-1.7}$\,cpd/100\,ton (68\% confidence interval), including systematic uncertainties. The inferred flux of CNO neutrino at Earth is $\Phi = 7.0^{+3.0}_{-2.0} \times 10^8 ~ \mathrm{cm}^{-2} ~ \mathrm{s}^{-1}$ (68\% confidence interval)~\cite{CNOpap}. The probability density function coming from the counting analysis is also reported. From the profiling of the log-likelihood, an exclusion of no-CNO hypothesis is achieved with 5.1$\sigma$ significance. A further hypothesis test with high statistics based on pseudo-datasets excludes the no-CNO hypothesis with more than 5.0$\sigma$ at 99\% confidence level.
The CNO neutrino measurement performed by Borexino is compatible with both HZ and LZ metallicity scenarios. A combined hypothesis test, including other solar neutrino fluxes measured by Borexino, shows a preference for the HZ hypothesis at 2.1$\sigma$ level, with still no significant improvement with respect to the Phase-II measurement.

\section{Conclusions}
After more than 13 years of data taking, Borexino has proven to be a unique instrument to study the Sun's interior. The unprecedented level of radiopurity achieved and the overall control of the detector backgrounds made possible the study of the basic processes which power our star, by looking at the neutrinos that are produced. The analysis of data collected during Phase-I and Phase-II provided the most accurate measurements of the \emph{pp} chain neutrino fluxes. In order to address the missing piece in the solar neutrino spectrum, the first ever detection of CNO neutrinos, a long-lasting effort for the detector stabilization was needed. The formation of a very clean region at the detector centre during Phase-III allowed to constrain the most important backgrounds for the CNO analysis and, in turn, to eventually detect the presence of CNO neutrino interactions in Borexino data. Thanks to Borexino, the two main processes which fuel the Sun's engine, and the one of all the other stars, have been finally disclosed. The precision on the measured neutrino flux, however, is still not precise enough to claim for a clear preference of one of the two metallicity scenarios.

\end{document}